\documentclass[10pt,showpacs,floatfix,letterpaper,amsmath,amsfonts,amssymb,aps,pra,reprint,superscriptaddress]{revtex4-1}
\usepackage[latin1]{inputenc}
\usepackage{bm}
\usepackage{graphicx}
\usepackage{tensor}
\usepackage{epstopdf}
\usepackage{tikz}
\usepackage[caption=false]{subfig}

\usetikzlibrary{shapes,positioning}
\tikzset{unit/.style={draw,shape=circle,fill=black,scale=0.2},costate/.style={draw,shape=trapezium},state/.style={draw,shape border rotate=180,shape=trapezium},qo/.style={draw,shape=rectangle},node distance=1mm,scalar/.style={draw,shape=regular polygon,regular polygon sides=6}}

\newcommand{\op}[1]{\hat{#1}}                                 
\newcommand{\rop}[1]{\check{#1}}                              
\newcommand{\qo}[1]{\mathcal{#1}}                             
\newcommand{\ket}[1]{\lvert #1\rangle}                        
\newcommand{\bra}[1]{\langle #1 \rvert}                       
\newcommand{\abs}[1]{\left\lvert #1 \right\rvert}             
\newcommand{\mean}[1]{\langle #1 \rangle}                     
\newcommand{\Tr}[1]{\text{Tr}[#1]}                            
\newcommand{\Trs}[1]{\text{Tr}_S[#1]}                         
\newcommand{\Trd}[1]{\text{Tr}_D[#1]}                         

\begin{document}
\title{Certainty in Heisenberg's uncertainty principle: \\Revisiting definitions for estimation errors and disturbance}
\author{Justin Dressel}
\affiliation{Center for Emergent Matter Science, RIKEN, Saitama 351-0198, Japan}
\author{Franco Nori}
\affiliation{Center for Emergent Matter Science, RIKEN, Saitama 351-0198, Japan}
\affiliation{Physics Department, University of Michigan, Ann Arbor, Michigan 48109-1040, USA}

\date{\today}

\begin{abstract}
We revisit the definitions of error and disturbance recently used in error-disturbance inequalities derived by Ozawa and others by expressing them in the reduced system space.  The interpretation of the definitions as mean-squared deviations relies on an implicit assumption that is generally incompatible with the Bell-Kochen-Specker-Spekkens contextuality theorems, and which results in averaging the deviations over a non-positive-semidefinite joint quasiprobability distribution.  For unbiased measurements, the error admits a concrete interpretation as the dispersion in the estimation of the mean induced by the measurement ambiguity.  We demonstrate how to directly measure not only this dispersion but also every observable moment with the same experimental data, and thus demonstrate that perfect distributional estimations can have nonzero error according to this measure.  We conclude that the inequalities using these definitions do not capture the spirit of Heisenberg's eponymous inequality, but do indicate a qualitatively different relationship between dispersion and disturbance that is appropriate for ensembles being probed by all outcomes of an apparatus.  To reconnect with the discussion of Heisenberg, we suggest alternative definitions of error and disturbance that are intrinsic to a single apparatus outcome.  These definitions naturally involve the retrodictive and interdictive states for that outcome, and produce complementarity and error-disturbance inequalities that have the same form as the traditional Heisenberg relation.
\end{abstract}


\maketitle


One of the fundamental principles of quantum mechanics is the uncertainty principle, which places a restriction upon the degree to which one can constrain the likelihoods of future measurements made on a quantum system. The initial form of this principle was postulated by Heisenberg \cite{Heisenberg1927} and rigorously derived by Kennard \cite{Kennard1927}.  Heisenberg subsequently corrected his original idea and presented its refined form in a remarkable series of lectures \cite{Heisenberg1930}.  The modern form of this principle extends the work of Heisenberg by directly relating this restriction to the noncommutativity of the algebra of quantum observable operators.  The abstract algebraic generalization was first derived for pure quantum states by Weyl and Robertson \cite{Weyl1928,*Weyl1950,Robertson1929}, but equally applies to other mathematical contexts (e.g., Fourier analysis).

A precise statement of the uncertainty principle from the quantum perspective is the following: If an experimenter repeatedly prepares a system in a particular quantum state $\op{\rho}$ and subsequently measures two observables $\op{A}$ and $\op{B}$ of the system using \emph{independent} preparations, then the accumulated statistics of the measurement will display variances, $\sigma_A^2 = \mean{(\op{A} - \mean{\op{A}})^2}$ and $\sigma_B^2 = \mean{(\op{B} - \mean{\op{B}})^2}$, that must satisfy the inequality
\begin{align}\label{eq:heisenberg}
  \sigma_A\,\sigma_B \geq \abs{\frac{1}{2i}\mean{[\op{A},\op{B}]}},
\end{align}
where $\mean{\op{A}} = \Tr{\op{A}\,\op{\rho}}$ is the usual expectation value, and $[\op{A},\op{B}]=\op{A}\op{B}-\op{B}\op{A}$ is the commutator of the observable operators.  This inequality follows in turn from the stronger Cauchy-Schwartz inequality that must hold for any operators $\op{A}$ and $\op{B}$
\begin{align}\label{eq:schrodinger}
  \sigma_A^2\,\sigma_B^2 \geq \left[\mean{\op{A} * \op{B}} - \mean{\op{A}}\mean{\op{B}}\right]^2 + \left|\frac{1}{2i}\mean{[\op{A},\op{B}]}\right|^2,
\end{align}
and which was derived for a pure quantum state $\op{\rho} = \ket{\psi}\bra{\psi}$ by Schr\"{o}dinger \cite{Schrodinger1930}.  Here we write it in terms of the symmetric Jordan product \cite{Jordan1934} of the operators $\op{A} * \op{B} = (\op{A}\op{B} + \op{B}\op{A})/2$, which will appear later in our discussion.

Recently it has become clear that the inequality in Eq.~\eqref{eq:heisenberg} is commonly (and perhaps incorrectly) associated with three conceptually distinct statements:
\begin{enumerate}
  \item A preparation $\op{\rho}$ has intrinsic spreads in $\op{A}$ and $\op{B}$ such that Eq. \eqref{eq:heisenberg} is satisfied for independent measurements.
  \item Estimating both $\op{A}$ and $\op{B}$ simultaneously will show measured estimation errors that satisfy Eq. \eqref{eq:heisenberg}.
  \item Estimating $\op{A}$ will disturb subsequent estimations of $\op{B}$ such that the measured estimation error of $\op{A}$ and disturbance of $\op{B}$ satisfy Eq. \eqref{eq:heisenberg}.
\end{enumerate}
All three of these statements relate to the original discussion by Heisenberg in Ref.~\cite{Heisenberg1927}, but only the first statement corresponds to the derivation of Eq.~\eqref{eq:heisenberg}.  

The remaining two statements correspond to experimental situations that do not relate to Eq.~\eqref{eq:heisenberg} directly, and that contain undefined new concepts.  The second statement concerns the complementarity (in the sense of Bohr) between the simultaneous \emph{estimation errors} for observables $\op{A}$ and $\op{B}$. The third statement concerns the trade-off between the estimation error of $\op{A}$ and the resulting \emph{disturbance} detectable by a subsequent measurement of $\op{B}$.  

Finding appropriate definitions (and resulting inequalities) that describe these distinct concepts of estimation error and disturbance has had a long history, much of which has been carefully reviewed by Busch \emph{et al.} \cite{Busch2007a}.  This pursuit can be traced back to the work of Arthurs and Kelly \cite{Arthurs1965}, who first demonstrated that a na\"{i}ve application of Eq.~\eqref{eq:heisenberg} will not describe the complementarity between estimations of a particular preparation state performed by a particular measuring apparatus.  They used a definition of the estimation error that generalizes the notion of classical mean-squared error for individual measurements \cite{Ozawa1988,Ozawa2013} (and which we will analyze more carefully later in this paper). Many papers subsequently appeared that highlighted different special cases where inequalities formally mimicking Eq. \eqref{eq:heisenberg} were inadequate for certain classes of states or observables using similar definitions that built off the work of Arthurs and Kelly (e.g., \cite{Arthurs1988,Martens1990,Ishikawa1991,Braginski1992,Jaeger1995,Appleby1998,Appleby1998a,Muynck2000,Trifonov2001}).  

A paradigm shift occurred when Ozawa \cite{Ozawa2003} derived a \emph{universal} inequality that was valid for any preparation state and a generic coupling to a detector
\begin{align}\label{eq:ozawa}
  \epsilon_A\,\eta_B + \epsilon_A\, \sigma_B + \sigma_A\,\eta_B \geq \abs{\frac{1}{2i}\mean{[\op{A},\op{B}]}}
\end{align}
using formal definitions for the mean-squared error $\epsilon_A$ and disturbance $\eta_B$ also suggested by Appleby \cite{Appleby1998,Appleby1998a} and based directly on the work of Arthurs and Kelly \cite{Arthurs1965}.  Shortly thereafter, Hall and Ozawa independently derived a similar universal inequality for the complementarity of joint observable estimations \cite{Hall2004,Ozawa2004a}
\begin{align}\label{eq:hall}
  \epsilon_A\,\epsilon_B + \epsilon_A\, \sigma_B + \sigma_A\,\epsilon_B \geq \abs{\frac{1}{2i}\mean{[\op{A},\op{B}]}},
\end{align}
building off Ozawa's previous result \footnote{Notably, Hall's result uses the standard deviations of optimal estimations rather than the intrinsic preparation deviations $\sigma_A$ and $\sigma_B$, so is in fact a stronger inequality than the one derived by Ozawa.}.  

Due to the similarity between these inequalities and Eq.~\eqref{eq:heisenberg}, as well as their resemblance to the discussion in Ref.~\cite{Heisenberg1927}, Eqs.~\eqref{eq:ozawa} and \eqref{eq:hall} have since been marketed as \emph{corrections} to Heisenberg's eponymous uncertainty principle that properly address the distinct concepts of estimation error and disturbance.  Nevertheless, these inequalities are still derived from Eq.~\eqref{eq:heisenberg} in its role as a generic operator inequality by making judicious replacements of $\op{A}$ and $\op{B}$.  As such, the interpretation of these new inequalities as corrections to the uncertainty principle crucially depends upon the physical significance of the additional quantities $\epsilon_A$ and $\eta_B$, and whether their definitions adequately reflect the situation considered by Heisenberg.

The definitions of the estimation error $\epsilon_A$ and the disturbance $\eta_B$ used by the inequalities in Eqs.~\eqref{eq:ozawa} and \eqref{eq:hall} can be understood as quantizations of classical mean-squared error and disturbance definitions \cite{Ozawa2013}.  Classically, these definitions involve the second moment of the difference between each measured result and well-defined reference values; the reference values are considered to be ``correct'' values, so that the differences may be interpreted as the ``errors'' of each individual measurement.  In contrast, the quantized versions formally involve the second moment of a difference between joint system-detector operators in the Heisenberg picture.  There are generally no well-defined reference values for each individual quantum measurement, so these formal definitions are the closest quantized equivalents to the classical definitions.  

Due to the formal nature of these definitions, and in spite of their correspondence to classical definitions, there has been controversy regarding both the physical significance of $\epsilon_A$ and $\eta_B$ in a quantum setting, as well as the feasibility of experimentally measuring them.  Indeed, there have been several independent appeals to find alternative and operationally motivated definitions \cite{Hofmann2003,Werner2004,Koshino2005,Busch2007,Busch2007a,Watanabe2011,Watanabe2011b,Hofmann2012a,Busch2013,Busch2013b} that produce inequalities which faithfully reflect Heisenberg's original discussion as presented in Ref.~\cite{Heisenberg1927}, and that also have a form similar to the inequality in Eq.~\eqref{eq:heisenberg}.

To address this controversy, Ozawa proposed an indirect method to experimentally determine the mean-squared error and disturbance \cite{Ozawa2004}, which has recently been implemented by Erhart \emph{et al.} \cite{Erhart2012}, Sulyok \emph{et al.} \cite{Sulyok2013}, and Baek \emph{et al.} \cite{Baek2013} using neutron-optical setups.  This method involves the preparation and measurement of three distinct but related states in order to formally construct the mean-squared error and disturbance associated with the measurement of one of those states.

To improve upon this indirect method by removing the need for three related preparations, Lund and Wiseman \cite{Lund2010} independently proposed an alternative procedure for measuring the mean-squared error and disturbance that requires only a single preparation state.  By expressing the error and disturbance in terms of a joint Terletsky-Margenau-Hill \cite{Terletsky1937,Margenau1961} quasiprobability distribution, they can be related \cite{Johansen2004} to weak values \cite{Aharonov1988,Duck1989} that may be approximately measured by a separate weakly correlated detector \cite{Dressel2010}.  This alternative procedure was recently implemented by Rozema \emph{et al.} \cite{Rozema2012} and Kaneda \emph{et al.} \cite{Kaneda2013} using optical setups.

A similar quasiprobability technique was recently proposed by Weston \emph{et al.} \cite{Weston2013} to derive (and then experimentally test) a tighter universal inequality for complementarity
\begin{align}\label{eq:weston}
  \epsilon_A\, \frac{\sigma_B + \sigma_{B,\text{est}}}{2} + \epsilon_B\, \frac{\sigma_A + \sigma_{A,\text{est}}}{2} \geq \abs{\frac{1}{2i}\mean{[\op{A},\op{B}]}},
\end{align}
where $\sigma_{j,\text{est}}$ represents the measured standard deviation of the estimation.  This tighter inequality has since been supplemented by inequalities derived by Branciard \cite{Branciard2013,Branciard2013b}
\begin{align}\label{eq:branciard}
  \sigma_A^2\, \epsilon_B^2 + \epsilon_A^2\, \sigma_B^2 + 2\, \epsilon_A^2\, \epsilon_B^2\, \sqrt{\sigma_A^2\, \sigma_B^2 - C_{AB}^2} \geq C_{AB}^2,
\end{align}
where $C_{AB} = \abs{\mean{[\op{A},\op{B}]}/2i}$.  Substituting $\eta_B$ for $\epsilon_B$ in Eq.~\eqref{eq:branciard} produces the tighter error-disturbance inequality corresponding to Eq.~\eqref{eq:ozawa}.  The Branciard inequalities have also been recently tested experimentally by Kaneda \emph{et al.} \cite{Kaneda2013}.

In this paper we take a step back from this rapid progression to reassess the formal definitions of the estimation error $\epsilon$ and the measurement disturbance $\eta$ used by the inequalities in Eqs.~\eqref{eq:ozawa}--\eqref{eq:branciard}.  To be clear, we do not challenge these results, since they are assuredly important and thought-provoking in their own right.  Instead, we explore the question of whether these results faithfully correct the idea of the uncertainty principle as discussed by Heisenberg in Refs.~\cite{Heisenberg1927,Heisenberg1930}, or whether they provide a qualitatively different and supplementary understanding to that principle.  By examining the definitions used in these inequalities from an instrumental approach \cite{Davies1970,Ozawa1984,Dressel2010,Dressel2012b,Dressel2013b}, we illustrate several subtle and unsatisfactory features that seem to deviate from the intent of Heisenberg.  

Despite the obvious correspondence to the classical ideas of mean-squared error and disturbance, the interpretation of the formal definitions of $\epsilon^2$ and $\eta^2$ as ``mean-squared deviations'' relies on an implicit assumption that observables can be assigned definite values before they are measured.  This assumption is in direct violation of the Bell-Kochen-Specker-Spekkens contextuality theorems \cite{Bell1964,Bell1966,Kochen1967,Spekkens2005,Spekkens2008}, and mandates the use of non-positive-definite quasiprobability distributions for interpreting the definitions as average deviations \cite{Ozawa2005a}.  Unlike classical observables, quantum observables generally do not have well-defined values prior to measurement from which one can construct meaningful deviations for each measurement realization.  Heisenberg was careful to avoid this appeal to any definite reference values \cite{Heisenberg1930}.

In the special case of an unbiased measurement (i.e., a measurement that faithfully determines the mean for any initial state), the quantity $\epsilon$ can be given a realistic statistical interpretation, though not as a root-mean-squared deviation.  Instead, it is the added \emph{dispersion} of the mean of an ensemble of measurements that stems from the ambiguity of the estimation.  However, we emphasize that $\epsilon$ does not quantify the error in the estimation of the full distribution for the observable, but only the dispersion in the estimation of its mean.  Indeed, we show how all moments of the estimated observable (and thus its full distribution) may be determined in a single experimental run even when $\epsilon$ is nonzero.  If the detector happens to report the same value range as the measured observable, then one can loosely interpret such a dispersion as an average deviation of each estimation from any possible reference value of the observable, but this loose interpretation fails for more general detector outputs that do not directly correspond to the spectral range of the measured observable.  We also highlight the relation between the mean-squared disturbance $\eta$ and the average Lindblad perturbation to the state that is induced by the estimating apparatus, and show that $\eta$ does not quantify this measurable perturbation in a natural way.

To reconnect with the original discussion of Heisenberg, we suggest an alternative perspective on estimation error and disturbance that focuses on what can be inferred on average from a \emph{single} apparatus outcome, rather than what can be inferred on average from \emph{all} apparatus outcomes.  Using alternative definitions of the estimation error and disturbance founded on recent work on the retrodictive and interdictive states associated with a single apparatus outcome \cite{Dressel2013b}, we derive and generalize inequalities that were independently obtained by Hofmann \cite{Hofmann2003}.  These inequalities also have the form of Eq.~\eqref{eq:heisenberg}, but correspond to the complementarity and error-disturbance interpretations of the Heisenberg uncertainty relation as it relates to a single apparatus outcome.  This result complements similar results obtained by Busch \emph{et al.} \cite{Busch2013,Busch2013b} and Watanabe \emph{et al.} \cite{Watanabe2011,Watanabe2011b}, who also use different definitions for error and disturbance that are intrinsic to the apparatus.

This paper is organized as follows.  In Sec.~\ref{sec:measurement} we briefly review the instrumental formalism of indirect observable measurement to keep the discussion self-contained.  In Sec.~\ref{sec:error} we examine the definition of estimation error $\epsilon$ as used by the inequalities of Eqs.~\eqref{eq:ozawa}, \eqref{eq:hall}, \eqref{eq:weston}, and \eqref{eq:branciard}.  In Sec.~\ref{sec:disturb} we examine the corresponding definition of measurement disturbance $\eta$.  In Sec.~\ref{sec:alternate} we consider alternative definitions of estimation error and disturbance that depend only on a single apparatus outcome, and consequently produce an alternative set of inequalities that have the same form as the traditional Heisenberg relation.  We conclude in Sec.~\ref{sec:conclusion}.

\section{Indirect Measurement}\label{sec:measurement}
The universal inequalities in Eqs.~\eqref{eq:ozawa}--\eqref{eq:branciard} all pertain to indirect observable measurements made using all outcomes of a detecting apparatus.  We will find it useful in what follows to discuss these sorts of measurements using a \emph{quantum instrument} approach \cite{Davies1970,Ozawa1984} augmented by the \emph{contextual-values} formalism \cite{Dressel2010,Dressel2012b,Dressel2013b}.  For completeness, we now briefly review this approach in both the detector picture (where the system is traced out) and the system picture (where the detector is traced out).

\begin{figure}[t]
    \includegraphics{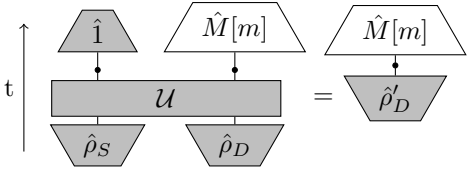}
    \caption{Detector picture of an indirect measurement, highlighted in gray, with time increasing along the vertical axis. (See \cite{Dressel2013b} for a more detailed description of the graphical notation used throughout this paper.) An initially uncorrelated system state $\op{\rho}_S$ and detector state $\op{\rho}_D$ are coupled with a unitary interaction $\qo{U}$. After the reference time indicated by the black dots, a joint observable $\op{1}\otimes\op{M}[m]$ with spectral function $m_k$ over the detector outcomes indexed by $k$ is projectively measured.  No further measurement is considered on the system space, which is indicated by the identity $\op{1}$ in the joint observable $\op{1}\otimes\op{M}[m]$.  This procedure is formally equivalent to the measurement of $\op{M}[m]$ directly after the preparation of a reduced detector state $\op{\rho}_D'$ that contains correlated information about the initial system state $\op{\rho}_S$.}
    \label{fig:indirectdetector}
\end{figure}

\subsection{Detector picture}
A system state---represented most generally by a density operator $\op{\rho}_S$---is coupled to an uncorrelated \footnote{The detector must be initially uncorrelated so it can be independently prepared for each trial and subsequently coupled to different system states.} detector state $\op{\rho}_D$ with an interaction $\qo{U}$ characterized by a unitary operator $\op{U}$, as illustrated schematically in Fig.~\ref{fig:indirectdetector}.  The joint state after the coupling is then correlated $\op{\rho}_{SD}' = \qo{U}(\op{\rho}_S\otimes\op{\rho}_D) = \op{U}(\op{\rho}_S\otimes\op{\rho}_D)\op{U}^\dagger$.  Performing a partial trace over the system produces the reduced detector state $\op{\rho}_D' = \Trs{\op{\rho}_{SD}'}$, which now contains information about the initial system state.  To exploit this correlation, the detector is read (i.e., measured projectively in some basis $\ket{k}$) and an eigenvalue $m_k$ is assigned by the experimenter to each detector outcome $k$.  After many identical trials, the average of the recorded values converges to the sum $\sum_k m_k\, p_k$, where $p_k = \bra{k}\op{\rho}_D'\ket{k}$ are the probabilities for observing the measured detector outcomes.

This empirical procedure corresponds to constructing a particular detector observable $\op{M}[m] = \sum_k m_k \ket{k}\bra{k}$, and measuring it projectively with respect to the reduced postinteraction detector state $\Trd{\op{M}[m]\,\op{\rho}_D'}$.  The chosen set of values $\{m_k\}$ assigned to each detector outcome $k$ forms the spectral function $m$ for the observable $\op{M}[m]$.  We keep the dependence of this observable on the chosen spectral function $m$ explicit, since its role in what follows will be important.

For brevity hereafter, we will use standard condensed notation for the joint expectation value $\mean{ \cdot } = \Tr{(\cdot)(\op{\rho}_S\otimes\op{\rho}_D)}$, the detector expectation value $\mean{ \cdot }_D = \Trd{(\cdot)\op{\rho}_D}$, the postinteraction detector expectation value $\mean{ \cdot }_{D'} = \Trd{(\cdot)\op{\rho}'_D}$, the system expectation value $\mean{\cdot}_S = \Trs{(\cdot)\op{\rho}_S}$, and the postinteraction system expectation value $\mean{ \cdot }_{S'} = \Trs{(\cdot)\op{\rho}'_S}$.

\subsection{System picture}\label{sec:system}
\begin{figure}[t]
    \includegraphics{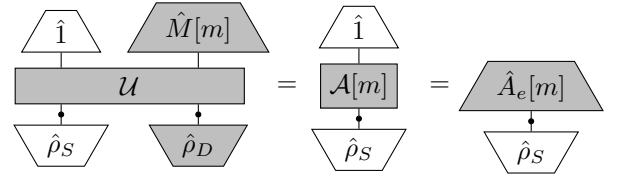}
    \caption{System picture of the same indirect measurement as in Fig.~\ref{fig:indirectdetector}, highlighted in gray.  The joint measurement procedure is equivalent to a \emph{quantum instrument} $\qo{A}[m]$ (see Sec.~\ref{sec:system}) that acts directly on the system after a reference time indicated by the black dots.  Since no subsequent measurements are considered (as indicated by the identity operator $\op{1}$), then this instrument induces a weighted POM $\op{A}_e[m]$ as the system observable that is effectively measured after the preparation of the initial state $\op{\rho}_S$, as shown in Eq.~\eqref{eq:equalities}.}
    \label{fig:indirectsystem}
\end{figure}

One can also perform a partial trace over the detector to produce a different picture of this empirical procedure that is contained entirely in the system space, as shown in Fig.~\ref{fig:indirectsystem}, and which will be more illuminating for what follows.  In this picture, observing a particular outcome $k$ on the detector induces a \emph{quantum operation} $\qo{A}_k$ on the system 
\begin{align}\label{eq:qo}
  \qo{A}_k(\op{\rho}_S) &= \bra{k}\op{U}(\op{\rho}_S\otimes\op{\rho}_D)\op{U}^\dagger\ket{k} \nonumber \\
  &= \sum_l p_l\, \bra{k}\op{U}\ket{l}\, \op{\rho}_S\, \bra{l}\op{U}^\dagger\ket{k} \nonumber \\
  &= \sum_l \op{M}_{k,l}\, \op{\rho}_S\, \op{M}_{k,l}^\dagger,
\end{align}
where each $\op{M}_{k,l} = \sqrt{p_l} \bra{k}\op{U}\ket{l}$ is a Kraus operator that characterizes the operation, and $\op{\rho}_D = \sum_l p_l\, \ket{l}\bra{l}$ is some (nonunique) pure state decomposition of the initial detector state.  It follows that the procedure for measuring the detector observable $\op{M}[m]$ equivalently weights a set of operations being performed on the system to produce a \emph{quantum instrument} 
\begin{align}
  \qo{A}[m] &= \sum_k m_k\, \qo{A}_k,
\end{align}
that completely describes the action of the indirect detector \cite{Davies1970,Ozawa1984,Ozawa2005,Dressel2013b}.  Intuitively, this instrument is the mathematical representation of the laboratory apparatus that is making the measurement.

Averaging the recorded values $m_k$ chosen by the experimenter therefore produces the set of formal equalities
\begin{align}\label{eq:equalities}
  \sum_k m_k \, p_k &= \mean{\op{U}^\dagger(\op{1}\otimes\op{M}[m])\op{U}} = \mean{\op{M}[m]}_{D'} \nonumber \\
  &= \Trs{\qo{A}[m](\op{\rho}_S)} = \mean{\op{A}_e[m]}_S
\end{align}
that are illustrated in Figs.~\ref{fig:indirectdetector} and \ref{fig:indirectsystem}.  The first equality is the joint system-detector picture that includes the interaction $\op{U}$ and the joint observable $\op{1}\otimes\op{M}[m]$.  The second equality is the detector picture involving the detector observable $\op{M}[m]$ and the reduced postinteraction detector state $\op{\rho}'_D$.  The third equality is the system picture involving the instrument $\qo{A}[m]$ acting on the initial system state $\op{\rho}_S$.  Since no further measurements are performed, the system picture can also be written as the last equality, which is the standard form for an expectation value of an effective system observable $\op{A}_e[m]$ in the initial system state $\op{\rho}_S$.

Due to the cyclic property of the trace, this effective observable can be understood as the action of the adjoint instrument $\qo{A}^*[m] = \sum_k m_k\, \qo{A}^*_k$ on the identity
\begin{align}\label{eq:cvs}
  \op{A}_e[m] &= \qo{A}^*[m](\op{1}),
\end{align}
where the adjoint instrument is composed of adjoint quantum operations
\begin{align}
  \qo{A}^*_k(\cdot) &= \mean{\op{U}^\dagger[ (\cdot)\otimes\ket{k}\bra{k}]\op{U}}_D = \sum_l \op{M}_{k,l}^\dagger(\cdot)\op{M}_{k,l}.
\end{align}
When acting on the identity, these adjoint operations induce a probability-operator measure (POM) \footnote{A POM also has the common name of positive-operator-valued measure (POVM).}
\begin{align}\label{eq:povm}
  \op{P}_k &= \qo{A}^*_k(\op{1}) = \sum_l \op{M}^\dagger_{k,l}\op{M}_{k,l}
\end{align}
over the outcomes $k$ of the detector such that $p_k = \mean{\op{P}_k}_S$ are the measured detector probabilities.  If all outcomes of the detector are accounted for, then the positive operators in the POM satisfy the normalization condition $\sum_k \op{P}_k = \op{1}$, making them a partition of unity.  In terms of this POM, the effective system observable has the intuitive form
\begin{align}\label{eq:cvspom}
  \op{A}_e[m] &= \sum_k m_k\, \op{P}_k.
\end{align}

\subsection{Measured observable}
The operator $\op{A}_e[m]$ in Eq.~\eqref{eq:cvspom} is the precise \emph{system} observable that is indirectly measured by the experimental procedure and chosen spectrum $m$.  Due to the operator equality in Eq.~\eqref{eq:cvspom}, the detector faithfully measures this observable with any initial system state $\op{\rho}_S$.  Note, however, that the experimenter-assigned values $m_k$ in the expansion of Eq.~\eqref{eq:cvs} are not generally the eigenvalues of $\op{A}_e[m]$.  Indeed, the number of detector outcomes $k$ may in fact be drastically different than the number of eigenvalues of $\op{A}_e[m]$. 

When the POM $\op{P}_k$ consists of projection operators then it follows that $\op{A}_e[m]$ and $\op{M}[m]$ have the same spectrum and the equality of Eq.~\eqref{eq:cvs} reduces to the spectral expansion of $\op{A}_e[m]$.  However, when the POM is not projective, then the function $m$ still constitutes a generalized spectrum for the measured observable $\op{A}_e[m]$ that corresponds to the specific induced POM $\op{P}_k$ for the measurement.  Such a generalized spectrum was dubbed a set of \emph{contextual values} for the observable $\op{A}_e[m]$ in previous work \cite{Dressel2010,Dressel2012b,Dressel2013b}, since the experimentally relevant set of values for an indirectly measured observable will depend on the context of exactly how it is being measured.

\section{Estimation Error}\label{sec:error}
Now suppose that we want to use an indirect measurement with instrument $\qo{A}[m]$ to estimate a particular system observable $\op{A} = \sum_a A_a \op{\Pi}_a$, with eigenvalues $A_a$ and spectral projectors $\op{\Pi}_a$.  How do we quantify the error of such an estimation?  To address this question, the inequalities of Eqs.~\eqref{eq:ozawa}--\eqref{eq:branciard} use the quantity
\begin{align}\label{eq:noise}
  \epsilon^2_A &= \mean{\op{N}^2},
\end{align}
which is the second moment of a joint \emph{noise operator} 
\begin{align}\label{eq:noiseop}
  \op{N} &= \op{U}^\dagger(\op{1}\otimes\op{M}[m])\op{U} - \op{A}\otimes\op{1}
\end{align}
under a specific initial joint state.  Ozawa demonstrated \cite{Ozawa2003,Ozawa2004,Ozawa2004a} that $\epsilon_A = 0$ if and only if the estimation is ``precise'' or ``noiseless,'' and that this definition reduces to the classical notion of mean-squared error when the two terms of Eq.~\eqref{eq:noiseop} commute \cite{Ozawa2013}.  Hence, this definition seems like a natural choice for quantizing the classical notion of mean-squared error.

In our notation, such a ``noiseless'' estimation implies the operator equality $\op{A} = \op{A}_e[m]$ between the effectively measured observable and the desired system observable.  Moreover, it implies that the induced POM $\op{P}_k$ must consist of the spectral projectors $\op{\Pi}_a$ of $\op{A}$, making it a projective measurement with the nonzero assigned values $m_k$ equal to the eigenvalues $A_a$.  In such a case, the number of nonzero $m_k$ must match the number of eigenvalues indexed by $a$, so that the detector spectrum and the observable spectrum are essentially identical.  Ozawa has argued that since the detector performs a projective measurement of the system observable in this case, then the system and detector are perfectly correlated and thus effectively ``have the same value(s)'' \cite{Ozawa2005a}, which justifies his terminology of ``noiseless'' estimation.

To better understand the quantity $\epsilon_A$, it is instructive to further dissect the noise operator $\op{N}$ in terms of the system picture with the detector traced out.  Observe that the first moment of the noise operator is 
\begin{align}\label{eq:difference}
  \delta_A &= \mean{\op{N}} = \mean{\op{A}_e[m] - \op{A}}_S,
\end{align}
which measures the difference in the estimated mean from the target value.  The second term involving the target value could be measured in a separate reference experiment, at least in principle.

In contrast, the second noise moment from Eq.~\eqref{eq:noise} simplifies to a less intuitive expression in the system space
\begin{align}\label{eq:difficultnoise}
  \epsilon^2_A &= \mean{\op{A}_e[m^2] + \op{A}^2 - 2 \op{A}_e[m]*\op{A}}_S
\end{align}
that contains the symmetric Jordan product [introduced in Eq.~\eqref{eq:schrodinger}] between the measured observable $\op{A}_e[m]$ and the target observable $\op{A}$.

The first term of Eq.~\eqref{eq:difficultnoise} is the second moment of the measured detector observable $\op{M}[m]$, which is equivalent to the effective system observable $\op{A}_e[m^2] = \sum_k m^2_k\,\op{P}_k$ obtained by squaring the spectrum $m$.  Note that $\op{A}_e[m]^2 \neq \op{A}_e[m^2]$ unless the POM is projective.  

The second term of Eq.~\eqref{eq:difficultnoise} is the second moment of the target observable $\op{A}$ in the initial system state.  As with the first moment in Eq.~\eqref{eq:difference}, this term could be measured in principle using a separate reference experiment.

The third term in Eq.~\eqref{eq:difficultnoise}, on the other hand, contains the Jordan product and does not obviously correspond to any measurable quantity pertaining to either the estimation experiment or a reference experiment.  Hence, we have a conundrum: Though the squared noise operator $\op{N}^2$ is formally a Hermitian observable that generalizes the notion of classical mean-squared error, it appears to be constructed from quantities that are not operationally meaningful for the situation under consideration.

To resolve this conundrum, Ozawa \cite{Ozawa2004} noted that it is possible to indirectly determine the problematic third term of Eq.~\eqref{eq:difficultnoise} if one is able to prepare not just one, but \emph{three} related system states: $\op{\rho}_S$, $\op{A}\op{\rho}_S\op{A}$, and $(\op{1}+\op{A})\op{\rho}_S(\op{1}+\op{A})$.  This indirect procedure follows from the identity,
\begin{align}\label{eq:difficultthirdterm}
  2\mean{\op{A}_e[m] * \op{A}}_S &= \mean{(\op{1}+\op{A})\op{A}_e[m](\op{1}+\op{A})}_S \nonumber \\
  &\qquad - \mean{\op{A}_e[m]}_S - \mean{\op{A}\op{A}_e[m]\op{A}}_S.
\end{align}
The additional operators $\op{A}$ and $(\op{1}+\op{A})$ that modify the state $\op{\rho}$ in each term can be understood as the Kraus operators for an auxiliary preparation apparatus.  Hence, all three states may be prepared without knowing $\op{\rho}$ \emph{a priori}, at least in principle, so each term in Eq.~\eqref{eq:difficultthirdterm} can be measured in different reference experiments.  This indirect procedure has been subsequently verified by Erhart \emph{et al.} \cite{Erhart2012}, Sulyok \emph{et al.} \cite{Sulyok2013}, and Baek \emph{et al.} \cite{Baek2013}.

While this indirect procedure clarifies that the quantity $\epsilon^2_A$ can be experimentally determined, it leaves several remaining conundrums.  First, it is not yet clear that $\epsilon^2_A$ is still operationally meaningful for a single experiment, even if it can be indirectly determined in multiple experiments.  Second, it is not clear exactly how $\epsilon^2_A$ corresponds to a ``mean-squared error'' quantum mechanically, since that intuition was based on a classical analogy.

\subsection{Quasiprobability interpretation}\label{sec:quasi1}
To address these remaining conundrums, Lund and Wiseman \cite{Lund2010} noticed that the Jordan product of two observables can be given a restricted interpretation as a meaningful quantity in terms of quasiprobabilities.  To see this, we also interpret the first two terms in Eq.~\eqref{eq:difficultnoise} as Jordan products with an appropriate identity operator, and expand $\epsilon_A^2$ directly in terms of the experimentally assigned values $m_k$ and the eigenvalues $A_a$ 
\begin{align}\label{eq:quasinoise}
  \epsilon_A^2 &= \sum_{k,a}(A_a - m_k)^2 \, \tilde{p}(a,k).
\end{align}
The distribution weighting this squared difference is a joint Terletsky-Margenau-Hill \cite{Terletsky1937,Margenau1961} \emph{quasiprobability distribution} 
\begin{align}
  \tilde{p}(k,a) &= \mean{\op{\Pi}_a * \op{P}_k}_S,
\end{align}
which is the real part of the Dirac \cite{Dirac1945,Chaturvedi2006,Lundeen2012}, or standard-ordering distribution \cite{Mehta1964}.  It explicitly involves the symmetric Jordan product of the spectral projection operators $\op{\Pi}_a$ of $\op{A}$ and the measured POM $\op{P}_k$ \footnote{This expansion generalizes the one used in Ref.~\cite{Lund2010}, which assumes $\op{M}[m]$ must share the same spectrum with $\op{A}$.}.  We use the tilde notation to indicate the quasiprobabilistic nature of the distribution, which can have negative values.

It is now easy to see that if $\tilde{p}(a,k)$ were a true joint probability distribution, then Eq.~\eqref{eq:quasinoise} would indeed compute the proper mean-squared deviation between the assigned values $m_k$ of the estimation and the eigenvalues $A_a$ of the target observable.  However, $\tilde{p}(a,k)$ is a quasiprobability distribution that is not positive definite unless $\op{A}_e[m]$ and $\op{A}$ commute with each other or the initial system state.  This lack of positivity is a manifestation of the Bell-Kochen-Specker-Spekkens contextuality theorems \cite{Bell1964,Bell1966,Kochen1967,Spekkens2005,Spekkens2008}.  Hence, Eq.~\eqref{eq:quasinoise} generally represents a mean-squared deviation only in a hypothetical, or \emph{counterfactual} sense.  It is the hypothetical deviation that \emph{would} correspond to an estimation error \emph{if} one could assign values to both observables $\op{A}$ and $\op{A}_e[m]$ simultaneously even without measuring them; however, the contextuality theorems prohibit exactly such a hypothetical joint value assignment, even when the observables commute.

\begin{figure}
    \includegraphics{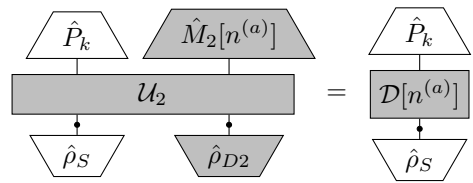}
    \caption{Weak measurement used to approximate the quasiprobabilities $\tilde{p}(a,k)$ according to Eqs.~\eqref{eq:condav} and \eqref{eq:jointav}, highlighted in gray.  A weak interaction $\qo{U}_2$ couples a second detector state $\op{\rho}_{D2}$ to the system, which is followed by a measurement of the detector observable $\op{M}_2[n^{(a)}]$ with assigned values $n^{(a)}_\ell$.  This joint procedure is equivalent to acting with a second quantum instrument $\qo{D}[n^{(a)}]$ on the system prior to measuring $\op{P}_k$, with operations $\qo{D}_\ell$ corresponding to each outcome $\ell$.  This detector is tuned to measure a spectral projection $\qo{D}^*[n^{(a)}](\op{1}) = \op{\Pi}_a$ of $\op{A}$ when the $\op{P}_k$ measurement is ignored.  However, subsequently measuring each $k$ results in the averages $\sum_\ell n^{(a)}_\ell p(\ell, k) \approx \tilde{p}(a,k)$ that approximate a quasiprobability distribution when $\qo{D}_\ell$ is nearly the identity for all $\ell$.  Changing the assigned values $n^{(a)}_\ell$ retargets the detector for different spectral projectors $\op{\Pi}_a$.}
    \label{fig:weakvalue}
\end{figure}

Nevertheless, these joint quasiprobabilities can still be approximately determined by introducing another detector into the experiment, as illustrated in Fig.~\ref{fig:weakvalue}.  To see this, we use Bayes' rule to split the joint quasiprobabilities into a product
\begin{align}\label{eq:quasiprob}
  \tilde{p}(k,a) &= \tilde{p}(a|k)\, p_k,
\end{align}
of the true detection probabilities $p_k = \mean{\op{P}_k}_S$ and \emph{conditional quasiprobabilities} \cite{Johansen2004}
\begin{align}\label{eq:weakvalue}
  \tilde{p}(a|k) &= \text{Re}\frac{\Trs{\op{P}_k \op{\Pi}_a \op{\rho}_S}}{\Trs{\op{P}_k\op{\rho}_S}}.
\end{align}
These conditional quasiprobabilities have the form of (real) \emph{generalized weak values} \cite{Aharonov1988,Duck1989,Wiseman2002,Dressel2010,Dressel2012b,Dressel2012d,Dressel2012e,Kofman2012,Dressel2013b,Dressel2013d} of the projection operators $\op{\Pi}_a$ with ``postselections'' corresponding to the measured POM $\op{P}_k$.  

These weak values can then be approximately measured according to Fig.~\ref{fig:weakvalue} by indirectly measuring the projection operators $\op{\Pi}_a$ using a second coupled detector \cite{Lund2010,Rozema2012,Weston2013,Kaneda2013}.  This second detector corresponds to an instrument $\qo{D}[n^{(a)}]$ with values $n^{(a)}_\ell$ assigned to its outcomes indexed by $\ell$.  When the original instrument $\qo{A}[m]$ is ignored, these values calibrate this new detector to measure a spectral projector of $\op{A}$ according to the identity $\op{\Pi}_a = \qo{D}^*[n^{(a)}](\op{1})$ analogous to Eq.~\eqref{eq:cvs}.  However, conditioning each outcome $\ell$ on a subsequently measured outcome $k$ of the instrument $\qo{A}[m]$ instead produces the conditioned average 
\begin{align}\label{eq:condav}
  \sum_\ell n^{(a)}_\ell p(\ell | k) &= \frac{\sum_\ell n^{(a)}_\ell p(\ell,k)}{\sum_\ell p(\ell,k)} 
\end{align}
in terms of the measured joint probabilities $p(\ell, k) = \Trs{\qo{A}_k(\qo{D}_\ell(\op{\rho}_S))} = \Trs{\op{P}_k \qo{D}_\ell(\op{\rho}_S)} = \mean{\qo{D}_\ell^*(\op{P}_k)}_S$.  This conditioned average approximates a weak value equal to the quasiprobability $\tilde{p}(a|k)$ when the detector operations $\qo{D}_\ell$ are sufficiently close to the identity operation for every outcome $\ell$ \cite{Dressel2010,Dressel2012b,Dressel2013b}.

We can also directly compute the \emph{measured} distribution that approximates the quasiprobability distribution of Eq.~\eqref{eq:quasiprob} without the appeal to intermediary weak values.  Since the measured $p_k$ in Eq.~\eqref{eq:quasiprob} is $\sum_\ell p(\ell, k)$, which is the denominator of the conditioned average in Eq.~\eqref{eq:condav}, we immediately obtain
\begin{align}\label{eq:jointav}
  \sum_\ell n^{(a)}_\ell p(\ell,k) &= \mean{\qo{D}^*[n^{(a)}](\op{P}_k)}_S \approx \tilde{p}(a,k).
\end{align}
This distribution is what was determined in the experiments by Rozema \emph{et al.} \cite{Rozema2012}, Weston \emph{et al.} \cite{Weston2013}, and Kaneda \emph{et al.} \cite{Kaneda2013}.

Introducing such an auxiliary weak detector necessarily perturbs the initial system state differently for each of the outcomes $\ell$, and thus modifies the experiment under consideration in a complicated way on average.  However, within some error tolerance each detection probability $p_k$ can be left approximately unaltered, and the joint distribution $\tilde{p}(a,k)$ can be approximately determined.  This solution, however, raises the question of why the definition of intrinsic estimation error of a detector requires the introduction of a second detector with its own estimation error and resulting disturbance on the initial state.  Determining the estimation error of that second detector would require a third detector, and so on.

\subsection{Unbiased measurements}
The interpretation of $\epsilon^2_A$ as a mean-squared error averaged with quasiprobabilities can be avoided, however, in the special case of an unbiased measurement, where one demands the estimated mean to be faithful (i.e., $\delta_A = 0$) for any initial state.  An operator equality $\op{A}_e[m] = \op{A}$ then follows from Eq.~\eqref{eq:difference} and the positivity of $\op{\rho}_S$ \cite{Dressel2010,Dressel2012b,Dressel2013b}.

Due to this operator equality, Eq.~\eqref{eq:difficultnoise} simplifies to
\begin{align}\label{eq:simplenoise}
  \epsilon^2_A &= \mean{\op{A}_e[m^2] - \op{A}^2}_S = \sum_k m_k^2\, p_k - \sum_a A_a^2\, p_a,
\end{align}
which is now completely analogous to the first moment $\delta_A$ in Eq.~\eqref{eq:difference}.  It is precisely the difference between the measured second moment using the raw detector values $m_k$ and the ideal second moment of $\op{A}$ that would be measured with its eigenvalues $A_a$ in a reference experiment.  The concreteness of this expression occurs because the quasiprobabilities $\tilde{p}(a,k)$ in Eq.~\eqref{eq:quasinoise} become positive-semidefinite when $\op{A}_e[m]$ and $\op{A}$ commute.

The second noise moment thus quantifies the amplification of the signal spread due to the weakened correlation with an unbiased detector.  Phrased in a different way, it quantifies the degree to which the assigned detector values have been amplified from the eigenvalues in order to compensate for the ambiguity of the measurement \cite{Dressel2010,Dressel2012b}.  It is now easy to see why $\epsilon_A = 0$ if and only if the measurement is projective: Only in that case will unbiased detector values match the eigenvalues for the observable.

Importantly, however, $\epsilon_A$ does not indicate the quality of the estimation of the observable \emph{distribution} that can be achieved using the experimental apparatus.  Indeed, the operator equality $\op{A}_e[m] = \op{A}$ guarantees that the first moment can be precisely obtained given a sufficiently large statistical ensemble of measurements.  Moreover, the same technique used to determine the spectral function $m$ to produce this equality can often be used to determine \emph{different} spectral functions $m^{(n)}$ that satisfy other equalities $\op{A}_e[m^{(n)}] = \op{A}^n$ for all powers of $\op{A}$ \cite{Dressel2010,Dressel2012b}.  In such a case, the same experimental data can be used to construct all higher moments of $\op{A}$ from the same measured set of probabilities, making the total experimental estimation of the distribution of $\op{A}$ essentially exact.  

To emphasize this point, consider the following diagonal two-outcome POM for a qubit:
\begin{align}\label{eq:pom}
  \op{P}_1 &= \begin{pmatrix}p_{11} & 0 \\ 0 & p_{12}\end{pmatrix}, & 
  \op{P}_2 &= \begin{pmatrix}p_{21} & 0 \\ 0 & p_{22}\end{pmatrix},
\end{align}
where $\op{P}_1 + \op{P}_2 = \op{1}$, and the $p_{ij}$ are positive probabilities such that $p_{i1}\neq p_{i2}$.  Examples of how to implement a POM of this sort can be found in Refs.~\cite{Pryde2005,Dressel2011,Goggin2011,Iinuma2011,Dressel2012c,Suzuki2012,Ota2012,Weston2013,Kaneda2013}.  It is straightforward to show \cite{Dressel2012b} that the POM in Eq.~\eqref{eq:pom} can be used to uniquely construct any diagonal operator $\op{A}$ according to
\begin{subequations}\label{eq:pomcvs}
\begin{align}
  \op{A} &= \begin{pmatrix}a_1 & 0 \\ 0 & a_2\end{pmatrix} = m_1 \op{P}_1 + m_2 \op{P}_2, \\
  m_1 &= \frac{p_{22}\, a_1 - p_{21}\, a_2}{p_{11}\,p_{22} - p_{21}\,p_{12}}, \\
  m_2 &= - \frac{p_{12}\, a_1 - p_{11}\, a_2}{p_{11}\,p_{22} - p_{21}\,p_{12}}.
\end{align}
\end{subequations}
Evidently the higher powers of $\op{A}$ are special cases of this solution that are obtained through the simple replacement $a_j \to (a_j)^n$ in the values $m_k^{(n)}$ for $\op{A}^n$.  Thus, all the moments of $\op{A}$ can be determined in \emph{one experimental run} of the detector by assigning appropriate values to each detector outcome in \emph{postprocessing}.  Notably, this method to determine the observable moments using a noisy signal is essentially classical in character since Eqs.~\eqref{eq:pom} and \eqref{eq:pomcvs} involve diagonal operators \cite{Dressel2012b}. 

It follows that one can construct $\op{A}^2 = \op{A}_e[m^{(2)}]$ with this technique, so one can compute $\mean{\op{A}^2}_S$ directly from the same data used to compute $\mean{\op{A}_e[m]}_S$.  Therefore, the second noise moment of Eq.~\eqref{eq:simplenoise} can be directly determined with no additional experimental work as the simple expression
\begin{align}\label{eq:simplenoisetwo}
  \epsilon^2_A &= \sum_k \left[ (m_k)^2 - m_k^{(2)} \right]\, p_k.
\end{align}
Hence, no reference experiments, quasiprobability arguments, or auxilliary detectors should be needed to directly determine this quantity in the laboratory, provided that one is using an unbiased detector.  

This demonstration also implies that the quantity $\epsilon_A$ in Eq.~\eqref{eq:noise} can be nonzero even when all moments of the measured observable can be faithfully determined by the same experimental data.  As such, it does not quantify the estimation error for the \emph{distribution} of the observable.  A better name for $\epsilon_A$ is the \emph{dispersion} of the estimation of the \emph{mean} of the observable, as also pointed out by Hall \cite{Hall2004}.  For an experimenter, a large dispersion corresponds directly to the need for an increased number of measurement realizations to produce good statistical results.  The more ambiguous (i.e., weak, or noisy) the measurement is, the larger the detector values must be, so the number of realizations needed to obtain the same statistical error for the mean will be larger.  Nevertheless, for a sufficiently large number of realizations even noisy measurements can be made statistically precise.

The interpretation of $\epsilon_A$ as a generic estimation error (as opposed to the dispersion of the mean) thus hinges crucially upon its decomposition into a mean-squared error for the \emph{individual measurement realizations}.  Since it does not indicate the estimation error for the full distribution of $\op{A}$, one can only argue that it provides a sensible notion of the average estimation error for each realization.  However, to adopt this point of view is to assert that the measured observable has a definite (correct) value prior to each measurement to which the (incorrect) detector result can be compared.  While this assertion is unproblematic for classical systems with definite values, it is a nontrivial assertion for a quantum mechanical system.  In contrast, interpreting $\epsilon_A$ as the dispersion in the estimation of the mean of the observable does not demand such a controversial statement.

\section{Measurement Disturbance}\label{sec:disturb}
\begin{figure}[t]
    \includegraphics{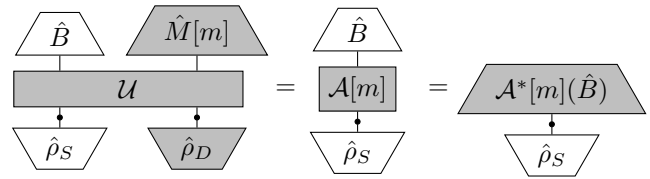}
    \caption{Sequential measurements.  After the indirect measurement illustrated in Fig.~\ref{fig:indirectsystem} and highlighted in gray, another system observable $\op{B}$ is measured.  The procedure produces correlated pairs of measurement outcomes $(m_k,B_b)$.  The quantum instrument $\qo{A}[m]$ is needed to fully describe these correlations in the reduced system space.  The effective system observable $\op{A}_e[m]$ from Fig.~\ref{fig:indirectsystem} reappears from the sequential observable $\qo{A}^*[m](\op{B})$ only when the correlations from the measurement of $\op{B}$ are ignored.}
    \label{fig:sequence}
\end{figure}

Now suppose we wish to measure a second observable $\op{B} = \sum_b B_b \op{\Pi}_b$ with eigenvalues $B_b$ and spectral projections $\op{\Pi}_b$ after the indirect measurement of $\qo{A}[m]$, as illustrated in Fig.~\ref{fig:sequence}.  To what degree has the act of measuring $\qo{A}[m]$ disturbed the subsequent measurement of $\op{B}$?  To address this question, the inequalities of Eqs.~\eqref{eq:ozawa} and \eqref{eq:branciard} use the quantity
\begin{align}\label{eq:disturb}
  \eta^2_B &= \mean{\op{D}^2},
\end{align}
which is analogous to Eq.~\eqref{eq:noise}, and is the second moment of a joint \emph{difference} operator 
\begin{align}\label{eq:diffop}
  \op{D} &= \op{U}^\dagger(\op{B}\otimes\op{1})\op{U} - \op{B}\otimes\op{1}.
\end{align}
between the original observable $\op{B}$ and the joint observable modified by the unitary interaction in the Heisenberg picture.  Ozawa demonstrated \cite{Ozawa2003,Ozawa2004,Ozawa2004a} that $\eta_B = 0$ if and only if the estimation of $\op{A}$ does not affect the subsequent measurement of $\op{B}$.  

\begin{figure}[t]
    \includegraphics{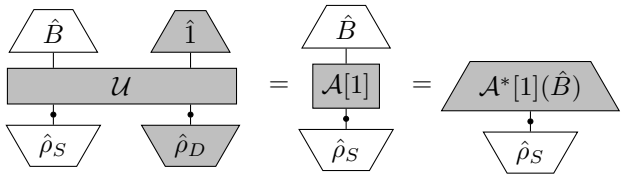}
    \caption{Disturbance induced by a nonselective measurement.  After the nonselective measurement highlighted in gray, another system observable $\op{B}$ is measured.  Unlike in Fig.~\ref{fig:sequence}, this procedure ignores the correlations between specific pairs of measurement outcomes, so will measure an average perturbed observable $\op{B}' = \qo{A}^*[1](\op{B})$.}
    \label{fig:nonselect}
\end{figure}

Again, it is instructive to dissect this definition in terms of the system picture with the detector traced out.  First note that the detector observable $\op{M}[m]$ with spectral function $m$ has been replaced by the identity $\op{M}[1] = \op{1}$ in the definition of Eq.~\eqref{eq:disturb}.  Setting the values $m_k$ to $1$ in this manner marginalizes over all the detector outcomes, which performs a \emph{nonselective} measurement using the entire apparatus $\qo{A}$.  This replacement implies that any information about the correlations between pairs $(m_k,B_b)$ of sequentially measured outcomes is being discarded by this procedure.  Only the net effect of averaging over \emph{all} outcomes $k$ is described by Eq.~\eqref{eq:disturb}, which is better illustrated by Fig.~\ref{fig:nonselect}.

We can write the first moment of the difference operator of Eq.~\eqref{eq:diffop} in several ways
\begin{align}\label{eq:bdifference}
  \delta_B &= \mean{\op{D}} = \Trs{\op{B}(\op{\rho}_S' - \op{\rho}_S)} = \mean{\op{B}' - \op{B}}_S,
\end{align}
where 
\begin{align}
  \op{\rho}_S' &= \qo{A}[1](\op{\rho}_S) = \sum_k \qo{A}_k(\op{\rho}_S) = \sum_{k,l} \op{M}_{k,l} \op{\rho}_S \op{M}_{k,l}^\dagger
\end{align}
is the reduced postinteraction state of the system, and
\begin{align}
  \op{B}' &= \qo{A}^*[1](\op{B}) = \sum_k \qo{A}^*_k(\op{B}) = \sum_{k,l} \op{M}_{k,l}^\dagger \op{B} \op{M}_{k,l}
\end{align}
is the Heisenberg operator that has been perturbed by the nonselective measurement.  Thus, the difference operator in Eq.~\eqref{eq:bdifference} provides information about how the mean of $\op{B}$ changes due to the evolution induced by the nonselective measurement.  

The squared disturbance in Eq.~\eqref{eq:disturb}, on the other hand, reduces to the expression
\begin{align}\label{eq:disturbsimple}
  \eta^2_B &= \mean{(\op{B}^2)' + \op{B}^2 - 2 \op{B}' * \op{B}}_S
\end{align}
on the system space, where 
\begin{align}
  (\op{B}^2)' = \qo{A}^*[1](\op{B}^2) = \sum_{k,l}\op{M}_{k,l}^\dagger \op{B}^2 \op{M}_{k,l}
\end{align}
is the squared Heisenberg operator perturbed by the nonselective measurement.  Analogously to Eq.~\eqref{eq:difficultnoise}, the expression of Eq.~\eqref{eq:disturbsimple} contains a symmetric Jordan product between the original observable $\op{B}$ and the perturbed observable $\op{B}'$, so is challenging to interpret in an experimentally meaningful way.  One can still indirectly determine the Jordan product term by exploiting the same identity used in Eq.~\eqref{eq:difficultthirdterm} of Sec.~\ref{sec:error}, however. 

\subsection{Quasiprobability interpretation}
As with the dispersion of Eq.~\eqref{eq:difficultnoise}, Lund and Wiseman \cite{Lund2010} observed that one can obtain an operational meaning for Eq.~\eqref{eq:disturbsimple} by expanding the Jordan product into a joint Terletsky-Margenau-Hill quasiprobability distribution.  To do so, we interpret the first two terms of Eq.~\eqref{eq:disturbsimple} as Jordan products with an appropriate identity operator, and expand $\eta^2_B$ directly in terms of the experimentally assigned values $B_{b'}$, the eigenvalues $B_b$,
\begin{align}\label{eq:quasidisturb}
  \eta^2_B &= \sum_{b,b'} (B_{b'} - B_b)^2\, \tilde{p}(b',b),
\end{align}
and the joint \emph{quasiprobability distribution} 
\begin{align}\label{eq:quasib}
  \tilde{p}(b',b) &= \mean{\op{Q}_{b'} * \op{\Pi}_b}_S.
\end{align}
that involves the Jordan product of the spectral projection operators of $\op{B} = \sum_{b} B_b\,\op{\Pi}_b$ and the perturbed POM $\op{Q}_{b'} = \qo{A}^*[1](\op{\Pi}_{b'})$ that constructs the perturbed observable $\op{B}' = \sum_{b'} B_{b'}\, \op{Q}_{b'}$ actually measured.

\begin{figure}
    \includegraphics{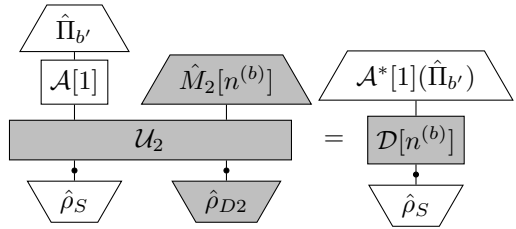}
    \caption{Weak measurement used to approximate the quasiprobabilities $\tilde{p}(b,b')$ in Eq.~\eqref{eq:quasib}, highlighted in gray.  A second quantum instrument $\qo{D}[n^{(b)}]$ is coupled to the system prior to the nonselective measurement $\qo{A}[1]$.  This detector is tuned to measure a spectral projection $\qo{D}^*[n^{(b)}](\op{1}) = \op{\Pi}_b$ of $\op{B}$ when all subsequent measurements are ignored.  However, measuring each $b'$ after $\qo{A}[1]$ results in the averages $\sum_{\ell,k} n^{(b)}_\ell p(\ell, k,b') \approx \tilde{p}(b,b')$ that approximate a quasiprobability distribution when $\qo{D}_\ell$ is nearly the identity for all $\ell$.  Changing the assigned values $n^{(b)}_\ell$ retargets the detector for different spectral projectors $\op{\Pi}_b$.}
    \label{fig:weakvalue2}
\end{figure}

As before, this joint quasiprobability distribution is not positive definite unless $\op{B}'$ and $\op{B}$ commute with each other or the initial system state.  Nevertheless, it can be approximately measured by probing the system weakly prior to the nonselective interaction of the apparatus $\qo{A}$ \cite{Lund2010,Rozema2012,Weston2013}, as illustrated in Fig.~\ref{fig:weakvalue2}.  This procedure is entirely analogous to the one discussed in Fig.~\ref{fig:weakvalue} and Sec.~\ref{sec:quasi1}.

Since determining a joint \emph{quasi}probability distribution is necessary in order to interpret Eq.~\eqref{eq:disturbsimple} as a mean-squared error caused by disturbance, the quantity $\eta^2_B$ does not generally pertain to any concrete notion of the disturbance inflicted upon $\op{B}$ by the apparatus $\qo{A}$.  Instead, $\eta^2_B$ pertains to a hypothetical, or \emph{counterfactual}, notion of disturbance.  That is, the observable $\op{B}$ is implicitly assigned a \emph{definite} value $B_b$ prior to the interaction for each realization.  That value is then disturbed to a different but equally definite value $B_{b'}$ with some transition quasiprobability $\tilde{p}(b',b)$ that averages over the effects of all intermediate $k$.  The non-positivity of the quasiprobability distribution indicates the questionable nature of this assumption that is incompatible with the Bell-Kochen-Specker-Spekkens contextuality theorems \cite{Bell1964,Bell1966,Kochen1967,Spekkens2005,Spekkens2008}.  

\subsection{Quantum nondemolition measurements}
If one demands that $\delta_B = 0$ for any initial state, then another operator equality $\op{B}' = \op{B}$ follows from Eq.~\eqref{eq:bdifference} and the positivity of $\op{\rho}_S$.  This equality is satisfied when $(\op{B}\otimes\op{1})$ commutes with $\op{U}$, or, equivalently, when $\op{B}$ commutes with $\op{M}_{k,l}$ for all $k,l$.  Notably, these commutation conditions are precisely the criteria for $\qo{A}$ to be a \emph{quantum nondemolition} (QND) measurement with respect to $\op{B}$ \cite{Nielsen2000}.

For such a QND measurement, the perturbed POM in Eq.~\eqref{eq:quasib} $\op{Q}_{b'} \to \op{\Pi}_{b'}$ reduces to the spectral projections of $\op{B}$ (up to relabeling of indices), the joint quasiprobability distribution becomes a true diagonal joint probability operator $\tilde{p}(b',b) \to p_b\, \delta_{b,b'}$, and the disturbance trivially vanishes: $\eta_B = 0$.  

\subsection{Lindblad decoherence}
The perturbation to $\op{B}$ can also be understood in terms of the induced \emph{Lindblad decoherence} stemming from the flow of system information to the discarded detector.  We can emphasize this connection by expanding the perturbed operator $\op{B}'$ using the following identity for each $k$:
\begin{align}\label{eq:sandwich}
  \sum_l \op{M}_{k,l}^\dagger \op{B} \op{M}_{k,l} &= \op{P}_k * \op{B} + \qo{L}_k(\op{B}),
\end{align}
where the probability operator $\op{P}_k = \sum_l\op{M}^\dagger_{k,l}\op{M}_{k,l}$ appears in a Jordan product, while the remainder $\qo{L}_k = \sum_l \qo{L}[\op{M}_{k,l}^\dagger]$ is composed of \emph{Lindblad operations} \cite{Lindblad1976} 
\begin{align}\label{eq:lindblad}
  \qo{L}[\op{M}^\dagger](\op{B}) &= -\frac{1}{2}\left(\op{M}^\dagger\,[\op{M},\op{B}] - [\op{M}^\dagger,\op{B}]\,\op{M}\right).
\end{align}
These operations were introduced in the study of decoherence for open quantum systems \cite{Breuer2007}.  In the present context the Lindblad operation indicates the average perturbation to $\op{B}$ that is induced by a particular measurement operator $\op{M}$.  

After summing the identity of Eq.~\eqref{eq:sandwich} over all $k$, we find the intuitive relation 
\begin{align}\label{eq:decoherence}
  \op{B}' - \op{B} &= \sum_k \qo{L}_k(\op{B})
\end{align}
between $\op{B}$ and its perturbation $\op{B}'$.  The difference in the measured mean $\delta_B$ from Eq.~\eqref{eq:bdifference} depends solely on this difference, so will be governed by the \emph{net} induced perturbation from the nonselective measurement. 

The same procedure can be applied to the second moment (and indeed all moments) to obtain the relation
\begin{align}\label{eq:decoherence2}
  (\op{B}^2)' - \op{B}^2 &= \sum_k \qo{L}_k(\op{B}^2).
\end{align}
We can thus insert the equalities of Eqs.~\eqref{eq:decoherence} and \eqref{eq:decoherence2} into Eq.~\eqref{eq:disturbsimple} to obtain
\begin{align}\label{eq:disturblindblad}
  \eta^2_B &= \sum_k \mean{\qo{L}_k(\op{B}^2) - 2 \op{B} * \qo{L}_k(\op{B})}_S.
\end{align}
Hence, only the Lindblad perturbation terms contribute to the squared disturbance $\eta^2_B$.  However, while both Eqs.~\eqref{eq:decoherence} and \eqref{eq:decoherence2} indicate measurable aspects of the disturbance to the moments of the distribution of $\op{B}$, the constructed quantity in Eq.~\eqref{eq:disturblindblad} still has an additional hypothetical character due to the remaining Jordan product.  In this sense, the manner in which the quantity $\eta_B$ quantifies the average perturbation to $\op{B}$ is unnatural from a distributional perspective. 

\section{An Alternative Perspective}\label{sec:alternate}
Heisenberg's original discussion in Ref.~\cite{Heisenberg1927}, on which the inequalities of Eqs.~\eqref{eq:ozawa}--\eqref{eq:branciard} are ostensibly based, was concerned with a different sort of estimation error and disturbance than we have been analyzing.  Rather than hypothetically tracking individual observable values being measured with all outcomes of an apparatus, Heisenberg considered how individual particles were actively affected by a \emph{single outcome} of the apparatus on average.  Depending on the ambiguity inherent to each apparatus outcome (e.g., a microscope with finite resolution for each measured interval), the future spreads of conjugate quantities would be altered in complementary ways on average.  This idea of average disturbance and estimation error corresponding to single apparatus events is not adequately captured by the definitions of $\epsilon_A$ and $\eta_B$, which rely on information obtained from all outcomes of the apparatus on average.  This point has also been made by Werner \cite{Werner2004}, Busch \emph{et al.} \cite{Busch2007,Busch2007a,Busch2013,Busch2013b}, Watanabe \emph{et al.} \cite{Watanabe2011,Watanabe2011b}, and Hofmann \cite{Hofmann2003,Hofmann2012a}.  

As such, we feel that the inequalities of Eqs.~\eqref{eq:ozawa}--\eqref{eq:branciard} do not capture the spirit of Heisenberg's original uncertainty discussion.  Instead, they are independently interesting and qualitatively different preparation-dependent measures of the dispersion of the estimation of the mean and the average quasiperturbation of individual eigenvalues.  We are thus led to consider alternative definitions of estimation error and disturbance that depend only on the individual instrument outcomes themselves.  This line of thought leads directly to several inequalities that have the same form as the Heisenberg relation in Eq.~\eqref{eq:heisenberg}, and which generalize the results in Ref.~\cite{Hofmann2003}.

\subsection{Estimation error}
\begin{figure}[t]
    \includegraphics{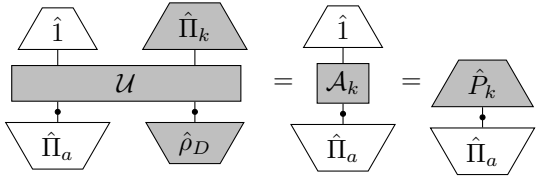}
    \caption{Operational procedure for determining the estimation error $\epsilon_{A,k}$.  For every preparation $\op{\Pi}_a$ in the spectral basis of $\op{A}$ made with probability $p(a)$, the system is prepared in the initial state $\op{\Pi}_a$ and subsequently conditioned on the successful measurement of a particular outcome $k$, which produces the measurable probabilities $p(k|a)$.  Applying Bayes' rule yields the retrodictive probabilities $p(a | k) = p(k | a)p(a) / \sum_a p(k|a)p(a)$.  For an uniform preparation with constant $p(a)$, this procedure produces $p(a|k) = \Trs{\op{\Pi}_a\,\rop{\rho}_k}$, which effectively normalizes the POM element $\op{P}_k$ to produce the retrodictive state $\rop{\rho}_k$ in Eq.~\eqref{eq:retrostate} that averages each possible preparation $\op{\Pi}_a$.  These probabilities define $\epsilon_{A,k}$ according to Eq.~\eqref{eq:erroralternate}.  The assumption of uniform $p(a)$ indicates that a single independent event $k$ cannot be assumed to correspond to any particular source preparation \emph{a priori}.  The resulting probabilities $p(a|k)$ are thus the best retrodictive inference about the preparation that one can make using only information about a single isolated detector outcome $k$.}
    \label{fig:errork}
\end{figure}

Since Heisenberg's arguments pertain directly to how individual outcomes of the measurement instrument affect any prepared particle on average, we examine the quantum instrument $\qo{A}[m]$ that is used to make the estimation without any reference to an initial system state.  Each outcome $k$ of this instrument produces a POM element $\op{P}_k$ according to Eq.~\eqref{eq:qo}.  As discussed in Refs.~\cite{Dressel2013b,Barnett2000a,Pegg2002a}, normalizing this POM element produces a \emph{retrodictive state} associated with the outcome $k$
\begin{align}\label{eq:retrostate}
  \rop{\rho}_k &= \frac{\op{P}_k}{\Trs{\op{P}_k}} = \frac{\sum_l \op{M}_{k,l}^\dagger\op{M}_{k,l}}{\sum_l \Trs{\op{M}_{k,l}^\dagger\op{M}_{k,l}}},
\end{align}
where we introduce an inverted hat to indicate the retrodictive nature of the state.

A natural notion of estimation error can then be defined as the \emph{retrodictive} standard deviation, $\epsilon_{A,k}$, which is the uncertainty in $\op{A}$ that can be retroactively inferred on average after obtaining the single outcome $k$ on the detector.  This is the best average uncertainty that one can infer with no prior information about the particle, and may be understood as the \emph{resolution} of the detector outcome $k$.  It is defined from the variance with respect to the retrodictive state for the outcome $k$
\begin{align}\label{eq:erroralternate}
  \epsilon_{A,k}^2 &= \sum_a (A_a - \mean{\op{A}}_k)^2 \, p(a | k) = \mean{\op{A}^2}_k - \mean{\op{A}}_k^2,
\end{align}
where $\mean{\cdot}_k = \Trs{(\cdot)\rop{\rho}_k}$ is the retrodictive expectation value, and $p(a | k) = \mean{\op{\Pi}_a}_k$ is the retrodictive probability of $a$ given the outcome $k$ \cite{Dressel2013b}.  This quantity can be directly measured as illustrated in Fig.~\ref{fig:errork} by comparing specific preparations of $\op{A}$ with resulting outcomes $k$.  Moreover, this quantity is a property of the specific instrument outcome $k$ without any reference to any particular preparation \cite{Amri2011}.

As an example, consider a common qubit POM $\op{P}_\pm = [\op{1} \pm \cos\theta\,\op{\sigma}_3]/2$ parametrized by an angle $\theta$.  This POM can be used to estimate the Pauli operator $\op{\sigma}_3$.  When $\theta = n\pi$ with integer $n$ then the POM is projective and the estimation error for each outcome should be zero.  Conversely, for $\theta = (n+1/2)\pi$ each POM element becomes the identity operator $\op{P}_\pm \to \op{1}/2$, so no information may be inferred about $\op{\sigma}_3$; therefore, the estimation error for each outcome should be maximal.  Computing each estimation error in Eq.~\eqref{eq:erroralternate} yields the constant $\epsilon_{\sigma_3,\pm} = \abs{\sin\theta}$, since the outcomes are symmetric.  This average error describes the resolution of each outcome and has the correct $\theta$ dependence.  Note that there is no \emph{a priori} fixed correspondence between the outcomes $\pm$ of the detector and the eigenvalues of $\op{\sigma}_3$. As such, $\epsilon_{\sigma_3,\pm}$ automatically chooses the optimal correspondence that has a maximum error of $1$ (as opposed to $2$ if the outcomes were permitted to correspond incorrectly to the eigenvalues in the projective case).

For comparison, the dispersion $\epsilon_{\sigma_3}$ used by Ozawa and others in Eq.~\eqref{eq:noise} depends not only on the entire POM, but also on the specific values $m_\pm$ that are assigned to each detector outcome.  Since $\epsilon_{\sigma_3}$ depends on this choice, it does not have a value that is intrinsic to the apparatus itself.  Assigning the values $m_\pm = \pm 1/\cos\theta$ to the detector makes the estimation unbiased, meaning $\sum_\pm m_\pm\, \op{P}_\pm = \op{\sigma}_3$; this is a special case of Eq.~\eqref{eq:pomcvs}.  If we pick these values, then Eq.~\eqref{eq:simplenoise} immediately implies that $\epsilon^2_{\sigma_3} = \sec^2\theta - \mean{\op{\sigma_3}}_S^2 \geq \sec^2\theta - 1 = \tan^2\theta$.  The second term is bounded by $1$ and depends on the preparation state, while the first term is state-agnostic and diverges as $\theta \to (n+1/2)\pi$.  The size of this dispersion is related to the number of realizations that are required to statistically determine the mean $\mean{\op{\sigma}_3}_S$ to an acceptable precision.  However, it does not directly indicate the resolution of each apparatus outcome, in contrast to our alternative definition in Eq.~\eqref{eq:erroralternate}.

Returning to the retrodictive estimation error, since Eq.~\eqref{eq:erroralternate} has the form of a standard deviation, the Weyl-Robertson inequality in Eq.~\eqref{eq:heisenberg} immediately implies the following inequality that generalizes one also derived by Hofmann \cite{Hofmann2003}
\begin{align}\label{eq:hofmann1}
  \epsilon_{A,k} \, \epsilon_{B,k} \geq \abs{\frac{1}{2i}\mean{[\op{A},\op{B}]}_k},
\end{align}
who considered outcomes with only a single measurement operator. The quantity $\epsilon_{B,k}$ is defined identically to $\epsilon_{A,k}$ in Eq.~\eqref{eq:erroralternate}, but with the substitution of the operator $\op{B}$. The bound involves the retrodictive expectation of the commutator between the two operators.

This inequality is a form of \emph{estimation complementarity} that holds specifically for a single outcome $k$ and agrees with the standard form of the Heisenberg uncertainty inequality.  It indicates that a single apparatus outcome $k$ cannot simultaneously estimate the values of two incompatible observables on average beyond a certain precision limited by an uncertainty bound.  As such, it corresponds to the second statement made in the introduction regarding complementarity, albeit with the important replacement of the preparation state with the retrodictive state corresponding to a specific instrument outcome $k$.  Indeed, note that for the canonical variables $\op{x}$ and $\op{p}$ considered by Heisenberg, this bound correctly reduces to the state-independent quantity $\hbar/2$.  Moreover, this inequality can also be trivially improved via the Cauchy-Schwartz form of Eq.~\eqref{eq:schrodinger}.

\subsection{Measurement disturbance}
\begin{figure}[t]
    \includegraphics{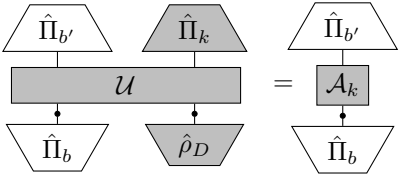}
    \caption{Operational procedure for determining the measurement disturbance $\eta_{B,k}$.  For each projection $\op{\Pi}_b$ in the spectral basis of $\op{B}$, the system in prepared in the initial state $\op{\Pi}_b$ with probability $p(b)$, the detector outcome $k$ is measured, then the outcome $\op{\Pi}_{b'}$ is measured, producing the measurable probabilities $p(k,b'|b)$.  Conditioning on the successful measurement of a particular intermediate outcome $k$ produces the interdictive joint probabilities according to Bayes' rule $p(b,b' | k) = p(k,b' | b)p(b) / \sum_{b,b'} p(k,b'|b)p(b)$.  When the preparation $p(b)$ is uniform, this yields $p(b,b'|k) = \Trs{\op{\Pi}_{b'}\,\tilde{\rho}_k(\op{\Pi}_b)}$, which effectively normalizes the operation $\qo{A}_k$ to produce the interdictive state operation $\tilde{\rho}_k$ in Eq.~\eqref{eq:interstate} that correlates each preparation $\op{\Pi}_b$ and post-selection $\op{\Pi}_{b'}$.  These probabilities define $\eta_{B,k}$ according to Eq.~\eqref{eq:disturbalternate}.  As with the procedure in Fig.~\ref{fig:errork}, the assumption of uniform $p(b)$ indicates the lack of \emph{a priori} bias that can be inferred by a single isolated outcome $k$.}
    \label{fig:disturbk}
\end{figure}

To describe an operational notion of disturbance, we need to consider correlations between measurable events prior and posterior to an intermediate outcome $k$.  As discussed in Ref.~\cite{Dressel2013b}, conditioning on an intermediate $k$ will normalize the corresponding transformation $\qo{A}_k$ in Eq.~\eqref{eq:qo}, which produces an \emph{interdictive state}
\begin{align}\label{eq:interstate}
  \tilde{\rho}_k(\cdot) &= \frac{\qo{A}_k(\cdot)}{\Trs{\qo{A}_k(\op{1})}} = \frac{\sum_l \op{M}_{k,l}(\cdot)\op{M}_{k,l}^\dagger}{\sum_l \Trs{\op{M}_{k,l}\op{M}_{k,l}^\dagger}}
\end{align}
that encodes all transformation information about that single detector outcome while leaving the choice of prior and posterior measurements unspecified.  This interdictive state is related to the retrodictive state for $k$ according to $\rop{\rho}_k = \tilde{\rho}^*_k(\op{1})$.  We use the tilde notation here to indicate that this exotic type of state is a \emph{transformation} and not simply an operator.

Using the interdictive state, we can naturally define an operational notion of disturbance as a mean-squared deviation between precise preparations and post-selections of an observable $\op{B}$ that bracket a specific outcome $k$ of the instrument
\begin{align}\label{eq:disturbalternate}
  \eta^2_{B,k} &= \sum_{b,b'} (B_b - B_{b'})^2 \, p(b,b' | k),
\end{align}
Unlike the quasiprobability distribution in Eq.~\eqref{eq:quasidisturb}, the joint distribution 
\begin{align}\label{eq:jointdisturb}
  p(b,b' | k) &= \Trs{\op{\Pi}_{b'}\tilde{\rho}_k(\op{\Pi}_b)} = \frac{\sum_l \abs{\bra{b'}\op{M}_{k,l}\ket{b}}^2}{\sum_l \Trs{\op{M}_{k,l}\op{M}^\dagger_{k,l}}}
\end{align}
is a true probability distribution, and directly corresponds to the experimental method illustrated in Fig.~\ref{fig:disturbk} for determining how each outcome $k$ disturbs definite preparations of $\op{B}$ on average.  This definition of disturbance produces an intrinsic property of the instrument that depends only on the outcome $k$ and the chosen observable $\op{B}$. 

Upon insertion of the Lindblad identity of Eq.~\eqref{eq:sandwich}, we can rewrite Eq.~\eqref{eq:disturbalternate} as
\begin{align}
  \eta^2_{B,k} &= \sum_{b,b'} (B_b - B_{b'})^2 \, \frac{\bra{b}\qo{L}_k(\op{\Pi}_{b'})\ket{b}}{\sum_l \Trs{\op{M}_{k,l}\op{M}_{k,l}^\dagger}}.
\end{align}
This form demonstrates that only the Lindblad perturbation $\qo{L}_k = \sum_l \qo{L}[\op{M}_{k,l}^\dagger]$ corresponding to the outcome $k$ will contribute to the mean-squared deviation $\tilde{\eta}^2_{B,k}$ in a natural way, in contrast to Eq.~\eqref{eq:disturblindblad}.

Our definition of disturbance in terms of an interdictive state reproduces and generalizes the one used in Ref.~\cite{Hofmann2003} for an outcome with a single measurement operator.  Hence, we can similarly define a restricted disturbance for each posterior $b'$ 
\begin{align}\label{eq:etatoepsilon}
  \eta^2_{B,k,b'} &= \sum_b (B_b - B_{b'})^2 \, p(b|k,b') \nonumber \\
  &= \epsilon^2_{B,k,b'} + (B_{b'} - \mean{\op{B}}_{k,b'})^2,
\end{align}
where $\mean{\op{B}}_{k,b'} = \Trs{\rop{\rho}_{k,b'}\op{B}}$ is the retrodictive mean under the state $\rop{\rho}_{k,b'} = \tilde{\rho}^*_k(\op{\Pi}_{b'})/p(b'|k)$ conditioned on both $k$ and $b'$.  For each $b'$ we then obtain the generalization to the second Hofmann inequality
\begin{align}\label{eq:hofmann2}
  \epsilon_{A,k,b'}\, \eta_{B,k,b'} \geq \epsilon_{A,k,b'}\, \epsilon_{B,k,b'},
\end{align}
which, combined with Eq.~\eqref{eq:hofmann1}, produces an estimation-disturbance inequality \emph{for each outcome pair} $(k,b')$ that agrees with the standard form of the Heisenberg uncertainty relation.  

Formally averaging Eq.~\eqref{eq:hofmann2} over every $b'$ with the probability distribution $p(b'|k) = \Trs{\tilde{\rho}^*_k(\op{\Pi}_{b'})}$ of obtaining $b'$ given $k$ produces the generalization of the third Hofmann inequality \cite{Hofmann2003}
\begin{align}\label{eq:hofmann3}
  \epsilon_{A,k}\, \eta_{B,k} \geq \abs{\frac{1}{2i}\mean{[\op{A},\op{B}]}_k}
\end{align}
that involves the full disturbance in Eq.~\eqref{eq:disturbalternate} and the retrodictive state for $k$ in a similar manner to the inequality of Eq.~\eqref{eq:hofmann1}.  The averaging step is a formal technique to derive the form of Eq.~\eqref{eq:hofmann3}, which holds generally.  Moreover, just as with Eq.~\eqref{eq:hofmann1}, the inequality of Eq.~\eqref{eq:hofmann3} may be further improved to match the form of Eq.~\eqref{eq:schrodinger} if desired.

The inequality of Eq.~\eqref{eq:hofmann3} corresponds to the third statement made in the introduction associated with the Heisenberg uncertainty relation that pertains to the error-disturbance trade-off intrinsic to a specific measurement outcome $k$.  As with the complementarity inequality, substituting the canonical operators $\op{x}$ and $\op{p}$ considered by Heisenberg produces the traditional state-independent bound of $\hbar/2$.

\subsection{Discussion}
To obtain results that are inconsistent with the inequalities of Eqs.~\eqref{eq:hofmann1} and \eqref{eq:hofmann3}, an experimenter must already know additional information about the initial state and combine that prior information with the new information obtained from the measurement.  This requirement of possessing information from multiple points in time has been recently emphasized by Rozema \emph{et al.} \cite{Rozema2013}, who defend the inequalities of Eqs.~\eqref{eq:ozawa}--\eqref{eq:branciard} as explicitly including the complete information about the initial state.  Such information is typically obtained from performing state tomography over an ensemble of different measurements made in separate experiments using the same preparation procedure.  This sort of estimation using general priors from multiple points in time necessarily applies to ensembles of measurements made by the detector, as we emphasized in the first half of this paper.

Heisenberg, however, was concerned with what one could infer on average from a single event of a detector at a local point in time, which is a qualitatively different scenario.  Indeed, according to \citet[][p. 20]{Heisenberg1930}, his principle
\begin{quote}
 ``\ldots states that every subsequent observation of the position will alter the momentum by an unknowable and undeterminable amount such that \emph{after} carrying out the experiment \emph{our knowledge} of the electronic motion is restricted by the uncertainty relation.''  (emphasis added)
\end{quote}
Indeed, any postmeasurement predictive state automatically satisfies the inequality \eqref{eq:heisenberg} concerning the intrinsic spreads of future measurements.  He goes on to clarify that \cite[][p. 20]{Heisenberg1930}
\begin{quote}
  ``\ldots the uncertainty relation does not refer to the past; if the velocity of the electron is at first known and the position then exactly measured, the position for times previous to the measurement may be calculated.  Then for these past times $\Delta p \Delta q$ is smaller than the usual limiting value, but this knowledge of the past is of a \emph{purely speculative character}, since it can never \ldots be subjected to experimental verification.  It is a matter of personal belief whether such a calculation concerning the past history of the electron can be ascribed any physical reality or not.'' (emphasis added)
\end{quote}

Evidently, Heisenberg thinks it is inappropriate to make an inference by combining the average knowledge about potential events at a prior point in time (i.e., the initial state) with detector events at a later point in time.  That combined knowledge can only be of a ``purely speculative character,'' and thus lead to \emph{counterfactual} characterizations of individual particle event chains, as we have pointed out with the definitions of error and disturbance used in the inequalities of Eqs.~\eqref{eq:ozawa}--\eqref{eq:branciard}.  This observation does not imply that these speculations about ensembles of past-particle histories are not interesting or important in their own right (e.g., \cite{Danan2013}), but it does challenge their interpretation as simple corrections to the uncertainty principle of Heisenberg.

Using only information that pertains to a single detector event at a particular time, on the other hand, does satisfy the spirit of Heisenberg's discussion of uncertainty.  As we have shown, making this restriction immediately produces the inequalities Eqs.~\eqref{eq:hofmann1} and \eqref{eq:hofmann3} for that isolated detector outcome, which complement the traditional Weyl-Robertson inequality in Eq.~\eqref{eq:heisenberg}.  Together, these inequalities correspond to all three interpretations of Heisenberg's uncertainty principle outlined in the Introduction, as it applies to each outcome of a detecting apparatus.  We also emphasize that other apparatus-intrinsic definitions of error and disturbance produce similar results \cite{Busch2013,Busch2013b,Watanabe2011,Watanabe2011b}, though we do not explore them here.

\section{Conclusion}\label{sec:conclusion}
In this work we have revisited the formal operator definitions of the mean-squared estimation error $\epsilon^2$ and disturbance $\eta^2$ used in recent inequalities derived by Ozawa and others that ostensibly generalize the Heisenberg uncertainty relation.  These important inequalities have recently been tested in \emph{tour de force} experiments.  Nevertheless, by analyzing the formal definitions in the system space, we have shown that they have unsatisfactory features that seem to deviate from the original discussion of Heisenberg.

Underlying the interpretation of the definitions of $\epsilon$ and $\eta$ as mean-squared deviations is the hidden assumption that observables have definite but unknown value assignments even when they are not measured.  This assumption is inconsistent with the Bell-Kochen-Specker-Spekkens contextuality theorems.  The nonpositivity of the quasiprobability distributions required to average the deviations indicates the questionable nature of the hidden assumption.  Moreover, these definitions pertain to inferences made from all outcomes of an apparatus on average, whereas Heisenberg considered what could be inferred from a single apparatus outcome on average.

For unbiased estimations, we clarified that the quantity $\epsilon$ can be given a concrete interpretation as the dispersion in the estimation of the mean.  Furthermore, we demonstrated how this quantity (and every observable moment) can be directly determined with the same experimental data used to estimate the mean, even when using non-projective measurements.  As such, $\epsilon$ does not quantify an estimation error for the full observable distribution; thus, its interpretation as an estimation error rests squarely upon its questionable decomposition into an average error of the individual measurement realizations themselves.

To reconnect with Heisenberg's original discussion, we considered alternative definitions of the estimation error and disturbance that focus on what one can infer on average from a \emph{single outcome} of an experimental apparatus.  By making the definition preparation agnostic, these alternative definitions describe properties intrinsic to the apparatus outcome itself.  The estimation error $\epsilon_k$ of a detector outcome $k$ indicates its intrinsic \emph{resolution}, and has the form of a standard deviation with respect to the \emph{retrodictive state} of the outcome.  This definition produces a complementarity inequality for each $k$.  These inequalities have the traditional form of the uncertainty relation, but substitute the retrodictive states for particular outcomes in place of the preparation state.

The analogous operational definition of disturbance $\eta_k$ for the outcome $k$ has the form of a root-mean-squared deviation between prior and posterior measurements conditioned on the intermediate outcome $k$.  This correlation can be written in terms of the recently introduced \emph{interdictive state} for the outcome.  This definition produces an error-disturbance inequality for each outcome $k$.  These inequalities also have the traditional form of the uncertainty relation, albeit with a similar substitution of the retrodictive states of each outcome.  

Therefore, the traditional Heisenberg uncertainty relation has been vindicated for single apparatus outcomes, even when applied to complementarity and error-disturbance arguments.

\begin{acknowledgments}
  JD would like to thank Curtis J. Broadbent, Adam Miranowicz, Abraham G. Kofman, Mark J. Everitt, and Eyob A. Sete for stimulating and enlightening discussions.  This work was partially supported by the ARO, RIKEN iTHES Project, MURI Center for Dynamic Magneto-Optics, JSPS-RFBR Contract No. 12-02-92100, Grant-in-Aid for Scientific Research (S), MEXT Kakenhi on Quantum Cybernetics, and the JSPS via its FIRST program.
\end{acknowledgments}

%
\end{document}